\documentclass[aps,pra,twocolumn,showpacs]{revtex4}
\usepackage{dcolumn}   
\usepackage{bm}
\usepackage{graphicx}
\usepackage{amsmath}   
\usepackage{latexsym}
\usepackage{amsfonts}   
\usepackage{amssymb}
\usepackage{array}     
\usepackage{epsfig}
\usepackage{epstopdf}
\usepackage{times}

\newcommand{\ket}[1]{\vert#1\rangle}
\newcommand{\miniket}[1]{\vert#1\rangle}

\newcommand{\ketbra}[2]{|#1\rangle \langle#2|}

\newcommand{\bra}[1]{\left\langle#1\right\vert}
\newcommand{\minibra}[1]{\langle#1\vert}

\newcommand{\valmed}[1]{\langle#1\rangle}

\begin{document}
\title{Characterizing multipartite symmetric Dicke states under the effects of noise}
\author{S. Campbell}
\author{M. S. Tame}
\author{M. Paternostro}
\affiliation{School of Mathematics and Physics, Queen's University, Belfast BT7 1NN, United Kingdom}

\begin{abstract}
We study genuine multipartite entanglement (GME) in a system of $n$ qubits prepared in symmetric Dicke states and subjected to the influences of noise. We provide general, setup-independent expressions for experimentally favorable tools such as fidelity- and collective spin-based entanglement witnesses, as well as entangled-class discriminators and multi-point correlation functions. Besides highlighting the effects of the environment on large qubit registers, we also discuss strategies for the robust detection of  GME. Our work provides techniques and results for the experimental communities interested in investigating and characterizing multipartite entangled states by introducing realistic milestones for setup design and associated predictions.
\end{abstract}
\date{\today}

\maketitle

\label{Intro}

The ability to classify entangled states~\cite{Wstates,class,class1} and quantify their degree of correlation~\cite{Horod} is advancing together with the capacity to experimentally produce interesting and useful forms of quantum correlated many-body systems. The range of experimentally available multipartite entangled states is witnessing a steady growth, stimulated by remarkable achievements such as the generation of eight-qubit W states~\cite{Wstates,guhneW}, ten-qubit GHZ-like states~\cite{ghz,guhnePan} and various sizes of cluster states~\cite{clusgraph,clusterExp}. Very successful linear optics experiments have been conducted exploring the many entangled classes of up to four-photon quantum correlated states~\cite{kiesel,schmidt,weinfurterDicke} and such possibilities are foreseeable in other physical systems as well~\cite{circuitQED}. Very recently, this range of interesting states has been enriched with the experimental generation of six-qubit symmetric Dicke states and their evaluation in multiparty quantum networking protocols~\cite{dicke,noidicke,munichdicke}, together with the exploration of a substantial part of the Dicke class of entangled states. Dicke states of $n$ qubits are defined as the eigenstates of both the total spin operator $\hat{J}^2=\sum_{k=x,y,z}\hat{J}^2_{k}$ and its $z$-component $\hat{J}_z$, where $\hat{J}_{k}=(1/2)\sum^n_{j=1}\hat{\sigma}^j_{k}$ and $\{\hat{\sigma}^j_{x},\hat{\sigma}^j_{y},\hat{\sigma}^j_{z}\}$ is the set of Pauli matrices of qubit $j$.
The model from which this class of states arise has been the focus of extensive investigations regarding super-radiance, quantum phase transition, correlation and entropic properties~\cite{variousVidal}. A Dicke state of $n$ qubits and $k$ excitations $|{D^{(k)}_n}\rangle$ is given by
\begin{equation}
|D^{(k)}_n\rangle = \frac{1}{\sqrt{C^k_{n}}}\sum_{l} \hat{P}_{l} \ket{0_{0} \dots 0_{k} 1_{k+1} \dots 1_{n}},
\end{equation}
where $C^k_{n}$ is the binomial coefficient and $\hat{P}_{l}$ is the set of all distinct permutations of $0$'s and $1$'s.  Symmetric Dicke states having $k=n/2$ are particularly interesting in virtue of the fact that they are associated with the largest eigenvalues of the set of observables $\{\hat{J}^2,\hat{J}_z\}$. Important theoretical studies~\cite{solano} and experimental progress in characterizing these states have been accompanied by the development and optimization of special tools explicitly designed in order to ``detect" the presence of genuine multipartite entanglement (GME) in a state. We define GME as states in which all subsystems are entangled with each other. In this context, a remarkable contribution has come from the introduction and use of {\it entanglement witness} operators (see~\cite{Horod,TothGuhneReview} and references within), which with often only modest experimental effort allow for the discrimination between separable, biseparable and fully GME states without requiring the complete knowledge of the state at hand. In fact, it is usually the case that  the complete detection and quantification of entanglement in a state requires knowledge of the full density matrix of the system. As this may not always be possible, entanglement witnesses provide us with viable ways to detect  entanglement through only partial information. Various forms of witnesses have been formulated recently, the most prominent being based on the use of state fidelity~\cite{NIELSEN} and collective spin-qubit operators~\cite{SS,vidal}.

It is very important to study the resilience of such characterization tools to the influences of unavoidable interactions between the constituents of a given system and their surrounding environment. These give rise to phenomena of dissipation and decoherence which have a negative impact on the entanglement content of a state. Knowing beforehand how a chosen method for GME-characterization is able to cope with such spoiling mechanisms is not only interesting, but also pragmatically useful. It allows one to make predictions about the performance of a setup and to determine in which direction technological progress should be made, in order to circumvent noise and reliably reveal quantum effects for fundamental studies and applications in quantum information tasks. Our work is performed precisely in this important direction. We concentrate on the class of symmetric Dicke states of $n$ qubits, which exhibit interesting quantum properties and are usable resources for quantum networking tasks such as quantum secret sharing, telecloning and open-destination teleportation~\cite{weinfurterDicke,noidicke}. We study the behavior of a variety of methods for revealing Dicke-class GME, including fidelity- and collective spin-based entanglement witnesses, as well as less-explored but valuable tools. We provide many setup-independent results that can be adapted to those experimental situations where local measurement settings can be reliably and easily arranged (as in Refs.~\cite{kiesel,circuitQED,solano}). This is a crucial requirement of the detection schemes addressed throughout this work. We also discuss feasible techniques for the improvement of the resilience of GME-detection to the influences of noise.

The remainder of the paper is organized as follows. Section~\ref{noise} briefly introduces the types of noisy channels considered in this study. Section~\ref{collectivebased} studies the behavior, under noisy mechanisms of collective-spin based entanglement witnesses of the form experimentally implemented in Refs.~\cite{noidicke,munichdicke} and proposes an original way to gain robustness. In Section~\ref{fidelitybased} we extend our investigation to more common fidelity-based entanglement witnesses. These are subjected to filtering operations in a way that stretches their tolerance to environmental effects and allows for a faithful detection of GME. We apply our techniques also to the class of $W$ states and highlight some unexpected differences among the channels. In Section~\ref{reducedbased}, the GME properties of symmetric Dicke states are highlighted by exploiting the interesting network of entanglement shared by reduced two-qubit states obtained by tracing out $n-2$ elements. Section~\ref{statediscri} approaches the problem of reliably discerning the entanglement class of a given noisy state. In Section~\ref{correlationbased} we explore ways to reveal the behavior of multi-qubit quantum coherence under the influence of noise, while Section~\ref{conclusions} briefly summarizes our findings. Finally, some technical details are presented in two appendices.

\section{Decoherence Models}
\label{noise}

In our study, we address a selection of physically relevant multi-qubit noisy channels affecting the class of states in question. Our choice encompasses a wide range of possible mechanisms that are likely to affect a given experimental setting designed to achieve multipartite entanglement. We use an effective picture for the action of a completely positive trace-preserving map given by the operator-sum representation~\cite{NIELSEN}. Within this formalism, a single-qubit noisy process is described by a set of Kraus operators $\{\hat{K}_\mu\}$, satisfying the completeness property $\sum_\mu\hat{K}^\dag_{\mu}\hat{K}_{\mu}=\hat{\openone}$, such that, calling $\rho_0$ the initial state of a qubit, its evolution is given by $\rho_{ch}=\$_{ch}(\rho_0)=\sum_\mu\hat{K}_\mu\rho_0\hat{K}_\mu^\dag$. In what follows, for the sake of convenience, $\gamma$ indicates the characteristic channel's influence rate, regardless of its specific nature and $ch$ is a label for the channel.

We start by considering the Kraus decomposition of a zero-temperature {\it amplitude damping} (AD) mechanism
\begin{equation}
\label{AD}
\hat{K}^{ad}_{0}=\ket{0}\bra{0}+e^{-\gamma/2}\ket{1}\bra{1},~~~~\hat{K}^{ad}_{1}=\sqrt{1-e^{-\gamma}}\ket{0}\bra{1}.
\end{equation}
This physically corresponds to an energy dissipation process: the system undergoing AD has a finite probability $e^{-\gamma}$ to lose an excitation (here, $\gamma$ is an effective dimensionless rate characterizing the whole process).   

The second process we consider is represented by pure {\it phase damping}  (PD) (or dephasing), which is a phase-scrambling and energy-preserving mechanism described by the two operators
\begin{equation}
\label{PD}
\hat{K}^{pd}_0=\sqrt{\frac{1+e^{-\gamma}}{2}}\hat{\openone},~~~~~\hat{K}^{pd}_1=\sqrt{\frac{1-e^{-\gamma}}{2}}\hat{\sigma}_z.
\end{equation}
It is easy to see that the action of PD on a single-qubit density matrix is to exponentially decrease the off-diagonal terms (at an effective rate $\gamma$), leaving the populations unaffected. Finally, we consider a {\it depolarizing} channel (DP), which (with probability $\gamma$) mixes a given {\it one-qubit} state with the maximally mixed state $\hat{\openone}/2$. Its action is given by the four-operator Kraus representation
\begin{equation}
\hat{K}^{dp}_{0}=\sqrt{1-\frac{3\gamma}{4}}\hat{\openone},~~\hat{K}^{dp}_{k}=\sqrt{\frac{\gamma}{4}}\hat{\sigma}_k~~~~(k=1,2,3),
\end{equation}
which gives $\rho_{dp}=(1-\gamma)\rho_0+\gamma\hat{\openone}_{}/2$. Therefore, the effect of a DP channel is to effectively add white noise to a given single-qubit state. This correspondence will become useful in our study. The above situation is different from that of a {\it collective} DP mechanism, which would add white noise to a multipartite state of $n$ qubits  $\rho_{mp}$, leading to $(1-\gamma)\rho_{mp}+\gamma\hat{\openone}_{}/2^n$.

When considering $n$-partite registers, the effects of equal noisy channels, each affecting an individual qubit of the system, can be accounted for by considering $q^n$ $n$-party tensor products of Kraus operators $\hat{\cal K}^{ch}_{j}$'s, where $q$ is the number of channel operators of a single-qubit channel.
That is
\begin{equation}
\rho_{ch}=\sum^{q^n}_{j=1}\hat{\cal K}^{ch}_j\rho_{mp}\hat{\cal K}^{ch\dag}_{j},
\end{equation}
Our task is now to provide an analysis as general as possible of the effects of such environmental channels on a variety of experimentally viable tools for multipartite entanglement.

\section{Collective-spin based entanglement witness}
\label{collectivebased}

Collective-spin operators are useful tools for the investigation of GME. Spin-squeezing inequalities fall into this class~\cite{SS} and have been extensively studied as well as experimentally implemented. More recently, collective-spin operators that are not directly related to spin squeezing have been formulated and shown to be particularly effective when symmetric, permutation invariant states are studied. One can construct the witness operator ~\cite{boundToth}
\begin{equation}
\label{standard}
\hat{\cal W}^s_{n}=b_{bs}\hat{\openone}-(\hat{J}^2_{x}+\hat{J}^2_{y}),
\end{equation}
where $b_{bs}$ is the maximum expectation value of $\hat{S}_n=\hat{J}^2_{x}+\hat{J}^2_{y}$ over the class of biseparable states of $n$ qubits. Finding $\langle\hat{\cal W}^s_n\rangle\!<\!0$ for a given state implies GME. The biseparable bound, $b_{bs}$ can be numerically calculated (see Appendix A and Ref.~\cite{boundToth}). The witness can be implemented with only two local measurement settings making it experimentally appealing and realizable in many physical settings (linear optics, circuit or cavity quantum-electodynamics). In particular, this tool has been used in~\cite{kiesel,munichdicke} for the case of four and six-qubit states.

Quite often, Eq.~(\ref{standard}) fails to detect GME in non-ideal symmetric Dicke states which have been affected by noise at their generation stage. Through a suitable modification, as experimentally demonstrated in~\cite{noidicke}, we can provide such a witness with greater flexibility in detecting GME in noise affected symmetric Dicke states. Let us introduce the generalized collective-spin witness
\begin{equation}
\label{betterone}
\hat{\cal W}^s_n(\alpha)=b_{bs}(\alpha)-\hat{S}_{n}(\alpha)~~(\alpha\in\mathbb{R}),
\end{equation}
where $\hat{S}_{n}(\alpha)=\hat{J}^2_{x}+\hat{J}^2_{y}+\alpha\hat{J}^2_{z}$. Here, we shall discuss how Eq.~(\ref{betterone}) offers more robustness to the noisy channels introduced in Sec.~\ref{noise} than the standard Eq.~(\ref{standard}). We notice that the bi-separability bound is now a function of new parameter $\alpha$. Using numerics it is seen that in general $b_{bs}(\alpha)\!<\!b_{bs}(0)$ for $\alpha<0$, which implies that at $\alpha\neq{0}$ the threshold for detection of GME is lowered. Consequently, we restrict our study to the case of negative $\alpha$.

Let us start with an AD channel and its effect on the state $\miniket{D^{(n/2)}_n}$. We call $\rho_{ad}$ the channel-affected version of this state, $\rho_{ad}=\$_{ad}(\miniket{D^{(n/2)}_n}\minibra{D^{(n/2)}_n})$, calculated as described in Sec.~\ref{noise}. One finds
\begin{equation}
\label{CollectiveAD}
\begin{aligned}
\text{Tr}[\hat{S}_n(\alpha)\rho_{ad}]\!&=\!\frac{n}{2}+\frac{n\alpha}{4}(n-1)(1-e^{-\gamma })^2\\
&+\frac{n^2}{4}e^{-\gamma}+\frac{n\alpha}{4}(1-e^{-2 \gamma}).
\end{aligned}
\end{equation}
\begin{figure}[t]
{\bf (a)}\hskip3cm{\bf (b)}
\psfig{figure=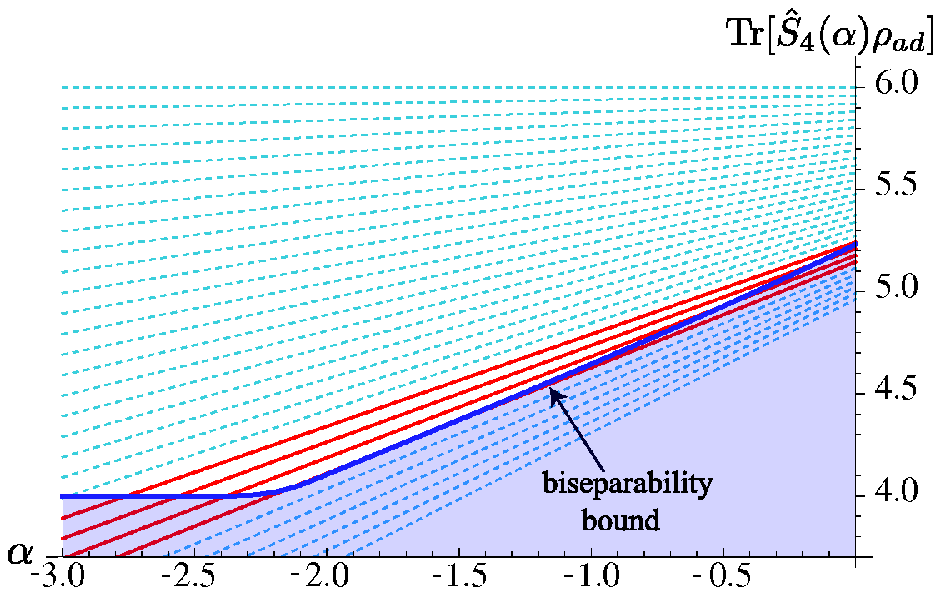,width=4cm,height=3.0cm}~~\psfig{figure=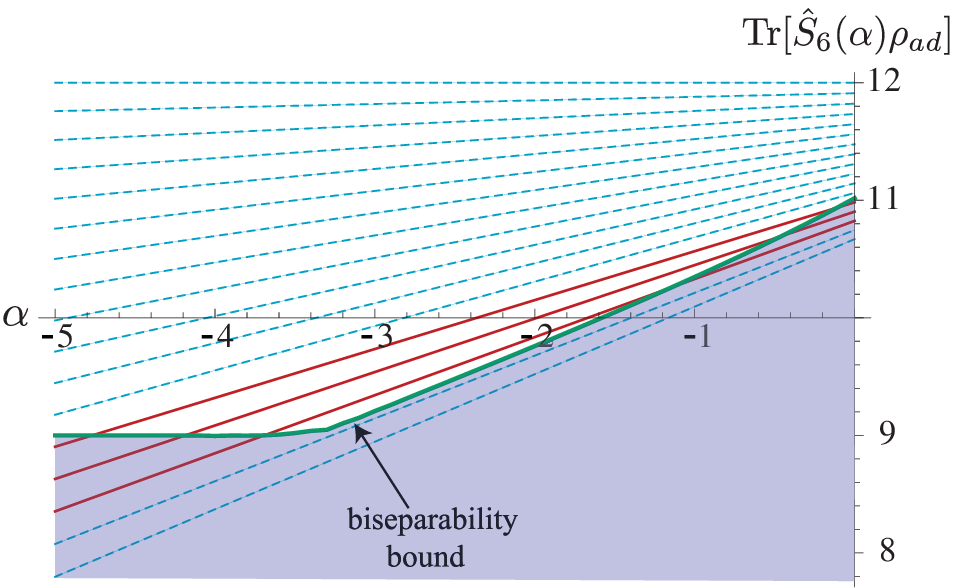,width=4cm,height=3.0cm}~~\caption{(Color Online) {\bf (a)} Collective-spin based entanglement witness $\langle\hat{S}_n(\alpha)\rangle$ for an AD-affected $\miniket{D^{(2)}_4}$ state. We show the lines corresponding to $\gamma\in[0,0.3]$ and highlight the cases where the modified entanglement witness turns out to be advantageous: each full line is such that $\langle\hat{\cal W}^s_4(0)\rangle>0$ with $\langle\hat{\cal W}^s_4(\alpha<0)\rangle<0$ in a region of values of $\alpha$. The behavior of the biseparability bound is also shown. {\bf (b)} Same as panel {\bf (a)} but for an AD-affected $\miniket{D^{(3)}_6}$ state. We show the lines corresponding to $\gamma\in[0,0.16]$. In both panels, the shading highlights the biseparability region.}
\label{figColl6}
\end{figure}
This formula is easily proven by noticing that, under a given noisy channel, the set of Pauli matrices of qubit $j$ changes as
$\hat{\sigma}^j_{k,ch}\!=\!\sum_{\mu}\hat{K}^{\dag}_{\mu}\hat{\sigma}^j_{k}\hat{K}^{}_{\mu}$. For AD, this leads to~\cite{carlo}
\begin{equation}
\label{carlo1}
\hat{\sigma}^j_{k,ad}=e^{-\gamma/2}\hat{\sigma}^j_{k},~~\hat{\sigma}^{j}_{z,ad}=(1-e^{-\gamma})\hat{\openone}+e^{-\gamma}\hat{\sigma}^j_z,
\end{equation}
where $k=x,y$. From these, it is easy to check that $\text{Tr}[\hat{\sigma}^{\otimes2}_{x,y}\otimes\hat{\openone}^{\otimes{(n-2)}}\rho_{ad}]\!=\!ne^{-\gamma}/(2n-2)$ while $\text{Tr}[\hat{\sigma}^{\otimes2}_{z}\otimes\hat{\openone}^{\otimes{(n-2)}}\rho_{ad}]\!=\!(1-e^{-\gamma})^2-e^{-\gamma}/(n-1)$. By using the explicit decomposition of $\hat{J}^2_{k}$ given in Appendix~A it is straightforward to see that
\begin{equation}
\label{demo}
\begin{aligned}
&\text{Tr}[(\hat{J}^2_x+\hat{J}^2_y)\rho_{ad}]\!=\!\frac{n}{4}(2+ne^{-\gamma}),\\
&\text{Tr}[\hat{J}^2_z\rho_{ad}]\!=\!\frac{1}{4} [n(n-1)(1-e^{-\gamma })^2+{n}(1-e^{-2 \gamma})],
\end{aligned}
\end{equation}
which leads directly to Eq.~(\ref{CollectiveAD}). We can infer two consequences. From the first identity of Eq.~(\ref{demo}) we see that for $\gamma\ge\ln[n^2/(4 b_{bs}-2n)]$, we have $b_{bs}(0)\ge\text{Tr}[(\hat{J}^2_x+\hat{J}^2_y)\rho_{ad}]$. This signals the failure of
Eq.~(\ref{standard}) and quantifies the amount of AD noise a given experimental set-up can tolerate. The second identity of Eq.~(\ref{demo}) shows that, different to a pure symmetric Dicke state, $\langle\hat{J}^2_z\rangle\neq{0}$ for $\rho_{ad}$ and strongly depends on $\gamma$. From Eq.~(\ref{CollectiveAD}) one finds that the dependence of the expectation value of $\hat{S}_n(\alpha)$ on $\alpha$ is linear, with a gradient coefficient determined by $\langle\hat{J}^2_z\rangle$, justifying the inclusion of such a term in the witness. For a decohered state such that  $b_{bs}(0)\ge\text{Tr}[(\hat{J}^2_x+\hat{J}^2_y)\rho_{ad}]$, the inclusion of the $z$-dependent term may be able to {\it pull} the expectation value of $\hat{\cal W}^s_n(\alpha)$ below zero, for a set value of $\alpha$. This is possible if $\langle\hat{J}^2_z\rangle$ is smaller than the gradient coefficient of the tangent to $b_{bs}(\alpha)$ at that point. In Fig.~\ref{figColl6} we show two instances of such a possibility for the cases of $n=4$ and $6$, for which $b_{bs}(0)\sim5.23$ and $11.018$ respectively (see Appendix~A and Refs.~\cite{noidicke,munichdicke}). Let us discuss the $n=6$ case: At $\gamma\ge0.116$ the witness in Eq.~(\ref{standard}) cannot detect GME in any AD-affected $\miniket{D^{(3)}_6}$. However, we find instances of decohered states that can be detected as GME via $\hat{\cal W}_6^s(\alpha<0)$, although $\gamma>0.116$. Such a possibility goes beyond the example given here. In fact, we have strong numerical evidence that for $\alpha=0$, GME cannot be revealed for $\gamma\ge{0.076}$ for an AD-affected $\miniket{D^{(4)}_8}$. However, for $\alpha\in[-5,-0.1]$ such states are detected as GME via our modified witness $\hat{\cal W}_6^s(\alpha)$ despite the fact that they correspond to $\gamma\in[0.18,0.20]$. Although a proof for any $n$ is difficult (due to the computational problem of quantifying $b_{bs}(\alpha)$ for large registers of qubits), this shows robustness of $\hat{\cal W}^s_n(\alpha)$ under the influences of AD channels.

\begin{figure}[t]
{\bf (a)}\hskip5cm{\bf (b)}
\psfig{figure=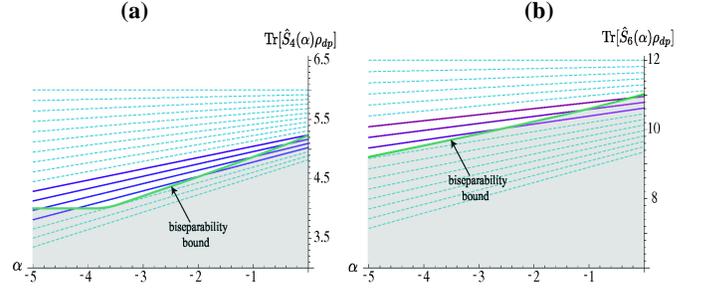,width=8.8cm,height=3.5cm}~~
\caption{(Color Online) {\bf (a)} $\langle\hat{S}_n(\alpha)\rangle$ for a DP-affected $\miniket{D^{(2)}_4}$ state. We show the lines corresponding to $\gamma\in[0,0.16]$ and highlight the cases where the modified entanglement witness turns out to be advantageous: each full line is such that $\langle\hat{\cal W}^s_4(0)\rangle>0$ with $\langle\hat{\cal W}^s_4(\alpha<0)\rangle<0$ in a region of values of $\alpha$. The behavior of the biseparability bound is also shown. {\bf (b)} Same as in {\bf (a)} but for $n=6.$ In both panels, the shading highlights the biseparability region.}
\label{figCollDep}
\end{figure}

A similar analysis conducted with respect to DP noise leads to the relation
\begin{equation}
\label{carlo2}
\hat{\sigma}^{j}_{k,dp}=(1-\gamma)\hat{\sigma}^j_k~~~(k=x,y,z),
\end{equation}
which tells us that the $\langle\hat{\sigma}^{\otimes{2}}_{x,y}\otimes\hat{\openone}^{\otimes(n-2)}\rangle$ are given by the expressions valid for a pure $\miniket{D^{(n/2)}_n}$ state, multiplied by $(1-\gamma)^2$. Following the same procedure as in the AD case leads us to
\begin{equation}
\label{demoDepo}
\begin{aligned}
&\text{Tr}[(\hat{J}^2_x+\hat{J}^2_y)\rho_{dp}]\!=\!\frac{n}{2}+\frac{n^2}{4}(1-\gamma)^2\\
&\text{Tr}[\hat{J}^2_z\rho_{dp}]\!=\!\frac{n}{4}\gamma(2-\gamma),
\end{aligned}
\end{equation}
from which the expectation value of the collective-spin based entanglement witness can be determined.
In this case too, the effectiveness of using $\hat{\cal W}_n^s(\alpha)$ is clearly revealed by the existence of situations where GME at non-zero $\alpha$ is detected, as shown in Fig.~\ref{figCollDep}, while this is not the case if the standard witness is used.

Finally, we address the PD channel, for which we have~\cite{carlo}
\begin{equation}
\label{carlo3}
\hat{\sigma}^{j}_{k,pd}=e^{-\gamma(1-\delta_{kz})}\hat{\sigma}^j_k~~~(k=x,y,z),
\end{equation}
where we have introduced the Kronecker-delta function $\delta_{kz}$ which is equal to $1$ for $k=z$ and $0$ otherwise. This clearly implies that a symmetric Dicke state affected by PD noise will keep the pure-state property $\text{Tr}[\hat{J}^2_z\rho_{pd}]=0$, for any $n$ and $\gamma$, therefore making the strategy highlighted so far ineffective. The expectation value of the witness operator reads, regardless of $\alpha$, as
\begin{equation}
\label{CollectivePD}
\text{Tr}[\hat{S}_n(\alpha)\rho_{pd}]\!=\!\frac{n}{2}+\frac{n^2}{4} e^{-2 \gamma},
\end{equation}
which is unable to detect GME as soon as $\gamma\ge\frac{1}{2}\ln[n^2/(4 b_{bs}-2n)]$. It remains to be checked whether, for instance, an approach such as the one used in Ref.~\cite{munichdicke,munichdicket} could be used in order to build up a resilient entanglement witness in this case as well.

\section{Fidelity-based entanglement witness and filtering}
\label{fidelitybased}
\subsection{Symmetric Dicke States}
\label{symdicke}

We now investigate fidelity-based witnesses~\cite{Horod,TothGuhneReview}. Despite generally requiring more local measurement settings than collective-spin based witnesses, they offer a more specific state characterization. Assuming that the state to study remains close to the Dicke class, the fidelity-based entanglement witness with the following general form can be used to detect GME
\begin{equation}
\label{wit}
\hat{\mathcal{W}}_n=c_n\hat{\openone}-\miniket{D^{(n/2)}_n}\minibra{D^{(n/2)}_n},
\end{equation}
where $c_n$ is the maximum overlap between $\miniket{D^{(n/2)}_n}$ and any possible biseparable state of $n$ qubits. Quantitatively, $c_n=n/(2n-2)$ for the class of symmetric Dicke states~\cite{TothGuhneReview,boundToth}.

When the AD-affected symmetric Dicke state is studied, based on our previous considerations, it is straightforward to find that
\begin{equation}
\label{fidwitnessAD}
\text{Tr}[{\hat{\cal W}_n\rho_{ad}}]=\frac{n}{2n-2}-e^{-\frac{n\gamma }{2}},
\end{equation}
regardless of $n$.
\begin{figure}[b]
{\bf (a)}\hskip3cm{\bf (b)}
\psfig{figure=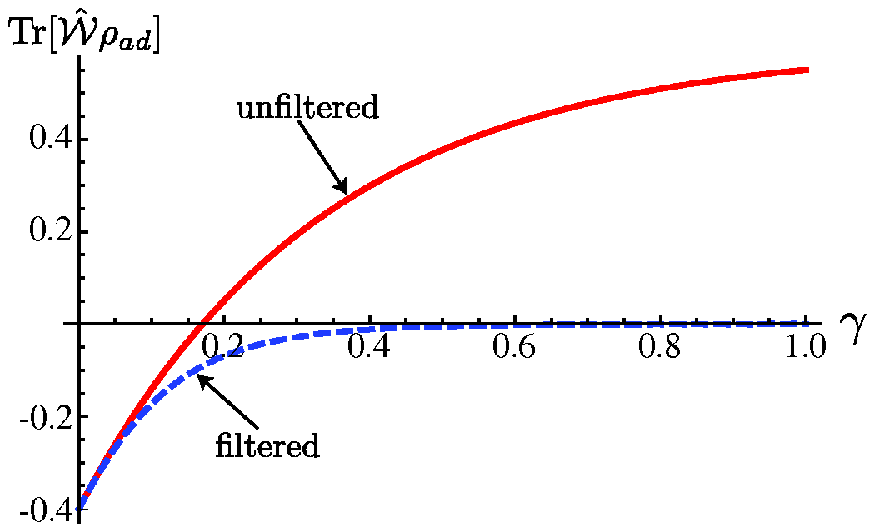,width=4cm,height=3cm}~~\psfig{figure=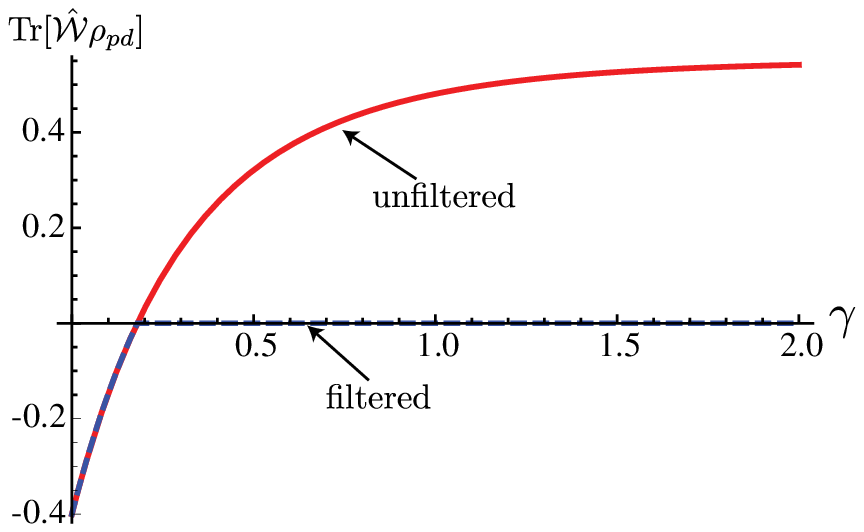,width=4cm,height=3cm}
\caption{(Color Online) {\bf (a)} Expectation value of an entanglement witness for a symmetric six-qubit Dicke state undergoing AD at a dimensionless rate $\gamma$. The solid curve corresponds to the unfiltered witness while the dashed curve is for the filtered one. By filtering one can increase the range of $\gamma$ where the witness is still able to detect GME. {\bf (b)} Expectation value of the entanglement witness for a symmetric six-qubit Dicke state undergoing PD type of noise. The solid curve is for the unfiltered witness while the dashed curve corresponds to the filtered case. No advantage is achieved, in this case, upon filtering.}
\label{fig1}
\end{figure}
For GME detection we require  $\text{Tr}[{\hat{\cal W}_n\rho_{ad}}]<0$, which is guaranteed for $\gamma<(2/n)\ln[(1/c_n)]$. For $n=4$ ($n=6$) the witness is implementable with $9$ ($21$) local measurement settings~(see Appendix B and Refs.~\cite{noidicke,munichdicke}). The limiting amount of sustainable AD noise for $n=4$ is $\gamma=0.203$ and decreases as larger Dicke states are examined (we have $\gamma=0.170$ for $n=6$, see Fig.~\ref{fig1} {\bf (a)}). Fig.~\ref{fig2} {\bf (a)} shows this ``pulling-back" effect for $n=4,..,50$. As previously done, we seek to devise experimentally-friendly techniques to detect GME for larger amounts of noise. To achieve this task, we apply {\it filtering operations}~\cite{toolbox,TothGuhneReview} to the fidelity-based entanglement witness. The local nature of the filters cannot alter the amount of entanglement in a state. However, they may allow for an increase in the noise allowed before a witness starts failing to detect GME. For $n$-qubit states, we thus construct filters of the form
\begin{equation}
\hat{\cal F}=\bigotimes^n_{j=1}\hat{\cal F}_{j}
\end{equation}
where $\hat{\cal F}_{j}$'s are local invertible operators. This technique has already been investigated for a variety of states in Refs.~\cite{Wstates,ghz,clusgraph,filter} and here we extend and generalize its use to arbitrarily-sized symmetric Dicke states.

We must be careful to ensure that the correct normalization is taken. To this end we impose the constraint Tr[$\hat{\mathcal{W}}_n$]=Tr[$\hat{\mathcal{W}}_n^{\mathcal{F}}$] with $\hat{\mathcal{W}}_n^{\mathcal{F}}$ the filtered witness operator. This is sufficient to ensure that the expectation value arising from the new witness is comparable to the expectation value from the unfiltered one. Therefore
\begin{equation}
\hat{\mathcal{W}}^{\mathcal{F}}_n=\frac{\textrm{Tr}[\hat{\mathcal{W}}_n]\hat{\mathcal{F}}\hat{\mathcal{W}}_n\hat{\mathcal{F}}^{\dagger}}{\textrm{Tr}[\hat{\mathcal{F}}\hat{\mathcal{W}}_n\hat{ \mathcal{F}}^{\dagger}]}.
\end{equation}
Filtering may lead to an increase in the number of required local measurement settings. However, there are indications that the class of filters given by~\cite{filter}
\begin{equation}
\hat{\mathcal{F}}_j=
\left(
\begin{array}{ll}
1 & 0 \\
0 & y_j  \\
\end{array}
\right)
\end{equation}
with $y_j$ a positive real number, guarantees for the smallest possible number of necessary measurement settings.
\begin{figure}[t!]
{\bf (a)}\hskip3cm{\bf (b)}
\psfig{figure=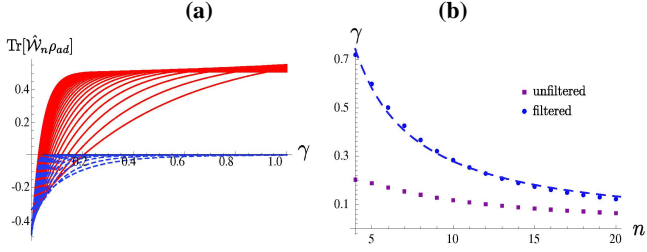,width=8.5cm,height=3cm}
\caption{(Color Online) {\bf (a)}  $\text{Tr}[\hat{\cal W}_n\rho_{ad}]$ plotted against the dimensionless AD rate $\gamma$ for $n=4\rightarrow{50}$. Solid (Dashed) lines are for unfiltered (filtered) witnesses. As $n$ grows, the filtering-induced gain is reduced. {\bf (b)} Values of $\gamma$ at which the filtered witness surpasses the threshold value $-10^{-3}$ and the unfiltered one starts to become non-negative with $n$. The lower (upper) curve is for the unfiltered (filtered) case. The advantage achieved upon filtering is clear as it slowly decays as $n$ increases. }
\label{fig2}
\end{figure}
A direct calculation, performed using (for convenience) $y_j=y~(\forall{j=1,..,n})$, leads to
\begin{equation}
\begin{split}
\text{Tr}[{\hat{\mathcal{W}}^{\mathcal{F}}_n\rho_{ad}}]&\!=\!\frac{(2^nn-2n+2) e^{-\frac{n\gamma }{2}}}{n \left(y^2+1\right)^n-(2n-2) y^n}\\
&\times[\frac{n}{2n-2} \left(y^2+e^{\gamma }-1\right)^{n/2}-y^n].
\end{split}
\end{equation}
This is then minimized with respect to the filtering parameter $y$ for any set value of $\gamma$, ensuring that the expectation value of the witness would not depend on $y$ at all~\cite{filter}. The effect of this procedure can be clearly seen in Fig.~\ref{fig1} {\bf (a)}: the local filtering increases the amount of decoherence tolerated while the witness is still able to detect GME in a symmetric Dicke state~\cite{footnote0}.

The actual detection of GME via negativity of a witness operator clearly depends on the errors associated with the experimentally determined value of $\text{Tr}[{\hat{\mathcal{W}}^{\mathcal{F}}_n\rho_{ad}}]$: a small (albeit negative) value is likely to be covered up by the corresponding error bar. Although the quantification of any acceptable lower bound for significant GME detection is a setup-dependent issue, based on current linear optics implementation, it is reasonable to expect that, upon collection of a sufficiently large sample of data, values of $\text{Tr}[{\hat{\mathcal{W}}^{\mathcal{F}}_n\rho_{ad}}]\sim-10^{-3}$ can still be discerned from zero. We thus fix a threshold of this order of magnitude and seek the smallest value of $\gamma$ at which the filtered witness (for a given $n$) surpasses it. This provides a practical lower limit to the mathematically rigorous performance of such tool. Fig.~\ref{fig2} {\bf (b)} reveals the advantage acquired upon filtering for $n=4\rightarrow{20}$: the gap with respect to the unfiltered case is quite considerable and marks the success of this strategy. Quantitatively, the upper points in Fig.~\ref{fig2} {\bf (b)} are well fitted by the function $\ln[(\frac{n-b_1}{n})^{\frac{2}{n}}{c_1}^{\frac{a_1}{n}}]$  with $a_1=1.54$, $b_1=-1.54$ and $c_1=4.73$.

We now study the detection of GME via a fidelity-based witness in symmetric Dicke states affected by PD type of noise and find
\begin{equation}
\label{fidwitPD}
\text{Tr}[{\hat{\mathcal{W}}_n\rho_{pd}}]=\frac{n}{2n-2}-\frac{1}{
C^{n/2}_{n}}
\sum_{k}^{{n}/{2}} (C^k_{n/2})^2
e^{-\gamma  (n-2 k)}.
\end{equation}
One can easily find that the general qualitative features highlighted for the case of the AD-related study of an unfiltered witness hold in this case as well: as $n$ grows, the values of $\gamma$ at which the expectation value of Eq.~(\ref{fidwitPD}) become positive are quickly pushed towards zero. However, an important remark is due: different to the case of AD noise, for PD the filtering technique employed above does not increase the range of tolerated noise. In fact, one can easily see that the net effect of the  application of filtering operators is that $\text{Tr}[{\hat{\mathcal{W}}^{\cal F}_n\rho_{pd}}]=y^n\text{Tr}[{\hat{\mathcal{W}}_n\rho_{pd}}]$. Therefore one cannot shift the value of $\gamma$ at which GME is unambiguously revealed. This is shown by the dashed line in Fig.~\ref{fig1} {\bf (b)}, where we see that the expectation value of the filtered operator tracks its unfiltered version in the negative semi-space~\cite{footnote}. The reason behind such a behavior is clearly understood by the noticing that,
\begin{equation}
\text{Tr}[{\hat{\mathcal{W}}^{\cal F}_n\rho_{pd}}]=\text{Tr}[\hat{\cal W}_n\sum^{2^n}_{j=0}\hat{\cal F}^\dag\hat{\cal K}^{pd}_j\miniket{D^{(n/2)}_n}\minibra{D^{(n/2)}_n}\hat{\cal K}^{pd\dag}_{j}\hat{\cal F}],
\end{equation}
where, as defined in Sec.~\ref{noise}, $\hat{\cal K}^{pd}_j$'s are the $n$-qubit Kraus operators for the whole PD-affected register. We can thus interpret this equation as the expectation value of the unfiltered witness over a symmetric Dicke state affected by new Kraus operators, each given by $\hat{\cal F}^\dag\hat{\cal K}^{pd}_j$. It is matter of direct calculation to see that the resulting density matrix $\tilde{\rho}_{pd}$ is simply $y^n\rho_{pd}$, thus demonstrating our claim. This is obviously not the case for an AD channel, in virtue of the different form of single-qubit Kraus operators required in that case. In order to exclude any limitations induced by the choice of the specific form of filtering operator, we repeated our analysis using general invertible operators, still finding no improvement upon filtering.

\begin{figure}[t]
{\bf (a)}\hskip3cm{\bf (b)}
\psfig{figure=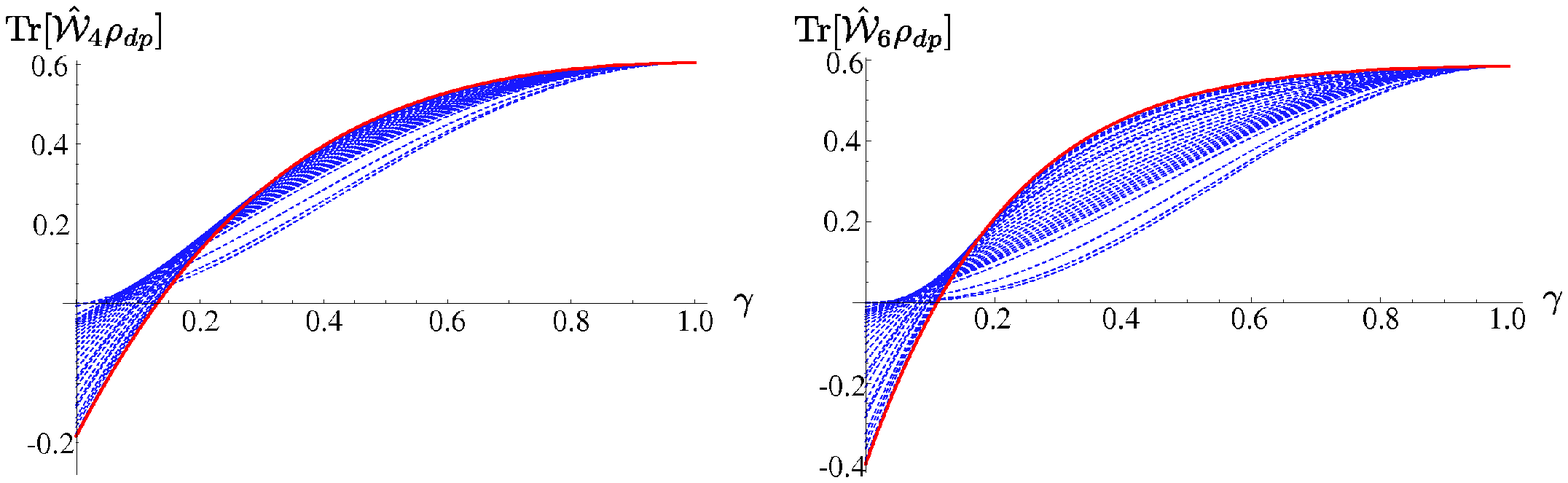,width=8.5cm,height=3.2cm}\\
\center{{\bf (c)}}\\
\psfig{figure=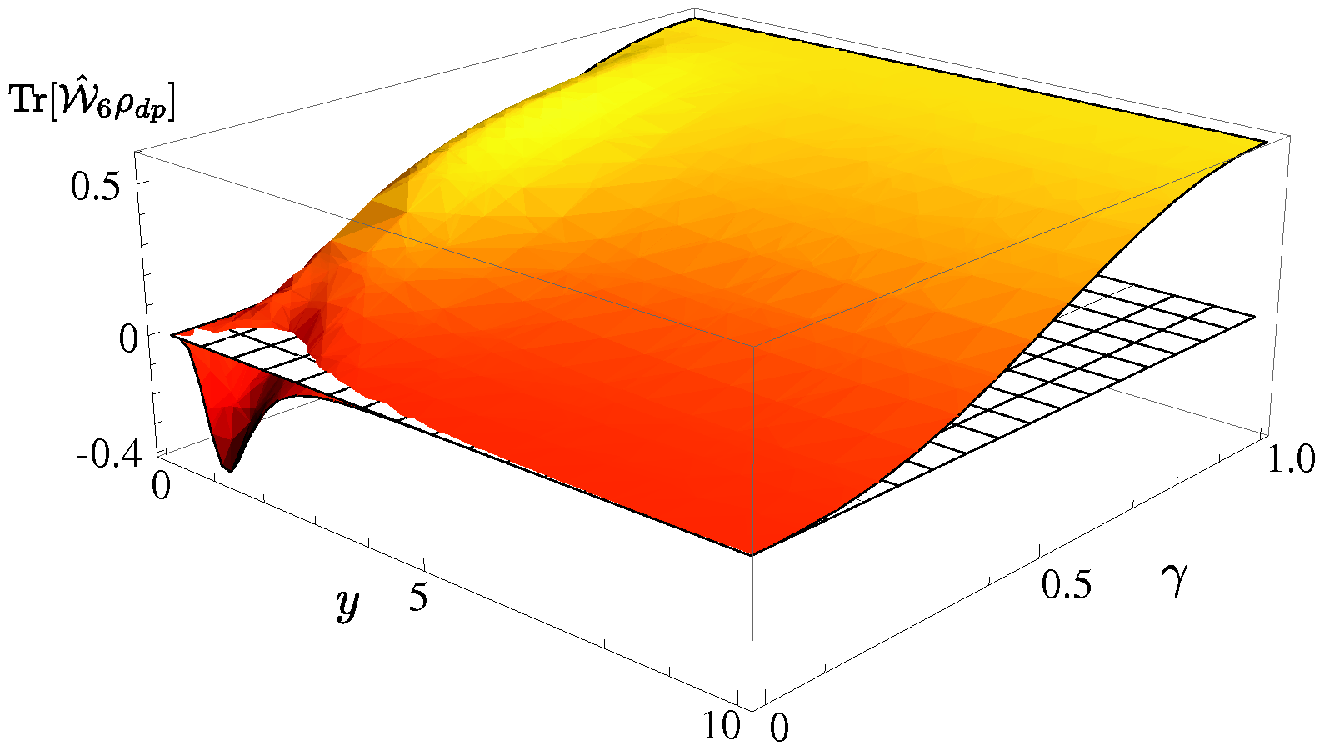,width=5cm,height=3.3cm}
\caption{(Color online) {\bf (a)} Expectation values of filtered and unfiltered fidelity-based entanglement witnesses for DP-affected $\miniket{D^{(2)}_4}$ plotted against the characteristic DP rate $\gamma$ (dimensionless). Each dashed line is for a filtered witness with $y\in[0,3]$ at steps of $0.1$. {\bf (b)} Same as in panel {\bf (a)} but for $\miniket{D^{(3)}_6}$. {\bf (c)} Filtered entanglement witness for DP affected $\miniket{D^{(3)}_6}$ against both $y$ and $\gamma$. In its negativity region, the witness is minimized at $y=1$.}
\label{figDepoWit}
\end{figure}

A very similar situation is encountered for the case of DP channels acting on the qubits of the register. In this case, the number of Kraus operators involved in the evolution of a given state grows as $4^n$. This makes any approach to the problem intractable for large $n$. Therefore, we have not been able to produce a general formula for the fidelity-based entanglement witness nor a simple argument to explain why, in this case as well, filtering is non-effective
in providing noise-robust GME detection. In Figs.~\ref{figDepoWit} {\bf (a)} and {\bf (b)} we give evidence of this for $n=4$ and $6$, comparing the unfiltered witness with a few instances of filtering. Moreover, panel {\bf (c)} shows that the filtered witness has an absolute minimum at $y=1$ in its region of negativity, demonstrating the ineffectiveness of filtering.

\subsection{W-States}
\label{Wstates}

With minimal changes to the analysis performed with respect to the symmetric Dicke states, we can also study $\ket{D^{(1)}_n}$, which are commonly referred to as $n$-qubit W states~\cite{Wstates}, thus showing the versatility of the techniques employed in this paper. We consider the fidelity-based witness~\cite{TothGuhneReview}
\begin{equation}
\label{witnessW}
\hat{\mathcal W}_w=\frac{n-1}{n} \openone-\miniket{D^{(1)}_n}\minibra{D^{(1)}_n},
\end{equation}
where in order to distinguish this case for the previously treated one, $\omega_{ch}$ is used to indicate the density matrix resulting from the application of channel $ch$ to a pure $n$-qubit W state.
Under AD, the expectation values of Eq.~(\ref{witnessW}) for the unfiltered and filtered cases are given by
\begin{equation}
\begin{aligned}
\text{Tr}[\hat{\cal W} _n\omega_{ad}]&=\frac{n-1}{n}\hat{\openone}-e^{- \gamma },\\
\text{Tr}[\hat{\cal W}^{\mathcal{F}}_n \omega_{ad}]&=\frac{[2^n (n-1)-n] [(n-1)(1-e^{-\gamma})-y^2{e}^{-\gamma}]}{n[(n-1) \left(y^2+1\right)^n-ny^2]}
\end{aligned}
\end{equation}
The effect of filtering once again increases the amount of tolerated noise for the witness to still be able to identify GME. Considering PD, we see that the addition of filtering has no benefit for the witnesses (for the same reasons explained in the discussion put forward in Sec.~\ref{symdicke}). We find the expectation value of the witness to be
\begin{equation}
\text{Tr}[\hat{\cal W}_n \omega_{pd}]=\frac{n-2}{n}-\frac{(n-1) e^{-2 \gamma }}{n}.
\end{equation}
Under appropriate conditions, these results are in agreement with the study reported for $n=4$ in~\cite{filter}. As for the DP channel, our investigation reveals that, at least for $n=4$ and $6$, a small advantage is gained by filtering the fidelity-based witness, although we do not have a closed analytical form for any $n$. 

\section{Reduced-State Based Entanglement Witness}
\label{reducedbased}

\begin{figure}[t]
{\bf (a)}\hskip3cm{\bf (b)}\\
\psfig{figure=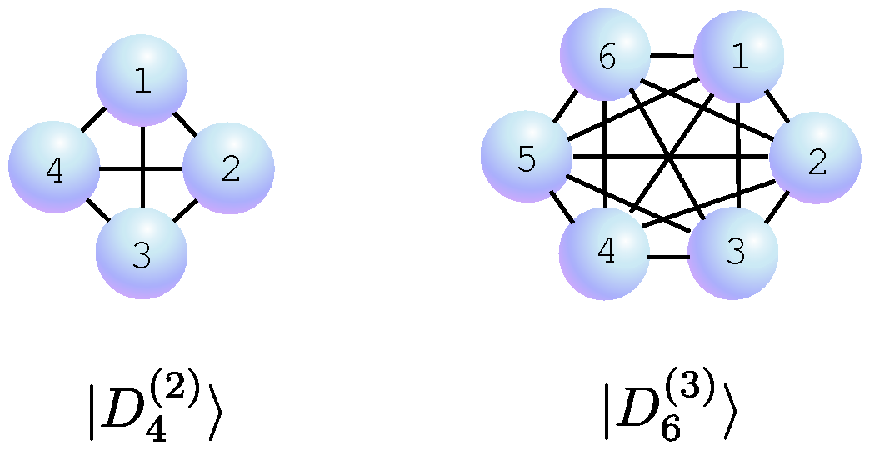,width=4cm,height=2.0cm} ~~\psfig{figure=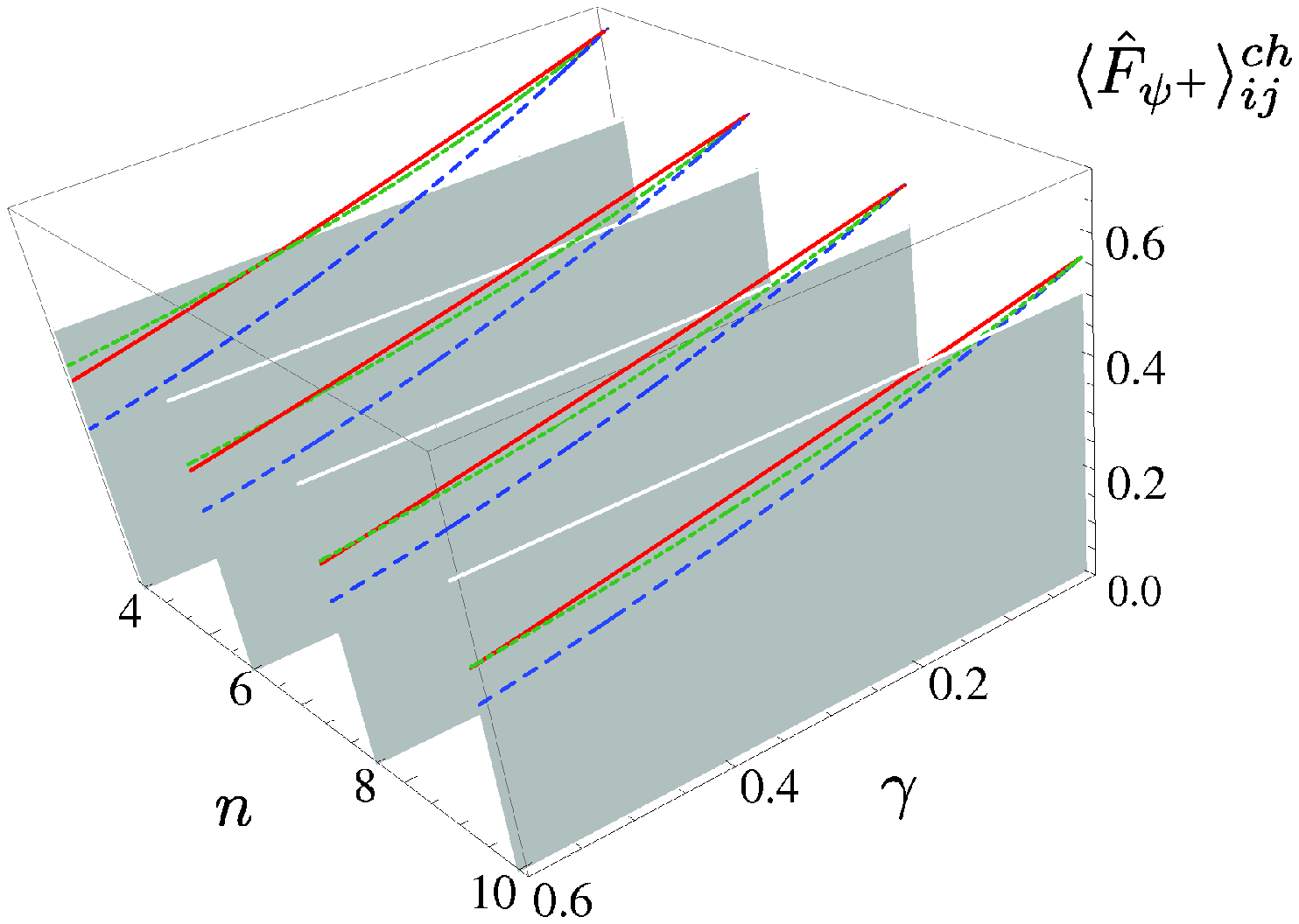,width=4cm,height=3cm}
\caption{
\textbf{(a)} Fully connected graphs for $\miniket{D^{(2)}_4}$ and $\miniket{D^{(3)}_6}$.
The vertices represent qubits and the edges represent the presence of entanglement within their reduced-state, as detected by using $\langle\hat{\cal W}_r \rangle$.
\textbf{(b)} Effects of noise on the detection of entanglement in the reduced two-qubit states of $|D_{n}^{(n/2)}\rangle$ using $\langle\hat{\cal W}_r \rangle$. Here, the solid (red), dashed (blue) and dotted (green) lines correspond to $\langle \hat{F}_{\psi^+}\rangle$ for the AD, DP and PD channels respectively. The shaded grey area corresponds to the region in which $\langle\hat{\cal W}_r \rangle$ fails to detect entanglement and the corresponding graph in \textbf{(a)} becomes completely disconnected.}
\label{figRed}
\end{figure}

We now consider a different manifestation of GME in symmetric Dicke states based on the observation of entanglement residing in the two-qubit reduced states of $|D_{n}^{(n/2)}\rangle$. Tracing out $n\!-\!2$
qubits, we find
\begin{equation}
\varrho=\alpha_n
\ketbra{\psi^+}{\psi^+}+\frac{(1-\alpha_n)}{2}[\ketbra{00}{00}+\ketbra{11}{11}]
\end{equation}
with $\alpha_n=n/[2(n-1)]$ for $n\ge4$~\cite{SS} and
$\ket{\psi^{\pm}}=(\ket{01}\pm\ket{10})/\sqrt{2}$.
Here, $\varrho$ has fidelity~\cite{NIELSEN} $\langle\hat{F}_{\psi^+}\rangle={\rm Tr}[\ketbra{\psi^+}{\psi^+}\varrho]\equiv\alpha_n$ with respect to $\ket{\psi^+}$.
Using the fidelity-based entanglement witness~\cite{TothGuhneReview}
\begin{equation}
\hat{\cal W}_r=\frac{1}{2}\openone-\ketbra{\psi^+}{\psi^+},
\end{equation}
we have $\langle\hat{\cal W}_r\rangle=1/2-\langle\hat{F}_{\psi^+}\rangle$. Thus, a fidelity $\langle\hat{F}_{\psi^+}\rangle>1/2$ detects the presence of entanglement in the two-qubit reduced state $\varrho$. This is always possible for all choices of pairs of qubits in $|D_{n}^{(n/2)}\rangle$, as $\alpha_n>1/2~\forall~n$. However, clearly, this may not be true for other states, including $|D_{n}^{(n/2)}\rangle$ subjected to the noise channels outlined in the Section~\ref{noise}. In general, we define two qubits of a multipartite state $\ket{\phi}$ as {\it connected} if their reduced density matrix is such that $\langle \hat{F}_{\psi^+}\rangle>1/2$. In this sense, a symmetric Dicke state gives rise to a connected set of reduced states.
For any given $n$, one can construct a graph having qubits $j=1,..,n$ at its vertices. Two vertices are joined by
an edge if and only if they are connected in the sense explained above. According to this definition, symmetric Dicke states give rise to fully connected (complete) graphs, as shown for $n=4$ and $6$ in Fig.~\ref{figRed}~{\bf (a)}.   

In order to relate the witness $\hat{\cal W}_r$ to observables in an experiment, we decompose the fidelity $\langle \hat{F}_{\psi^+}\rangle$ into expectation values of Pauli operators as follows
\begin{equation}
\langle \hat{F}_{\psi^+}\rangle\!=\!\frac{1}{4}(1+{\rm Tr}[\hat{\sigma}_x\otimes\hat{\sigma}_x\varrho]+{\rm Tr}[\hat{\sigma}_y \otimes\hat{\sigma}_y \varrho]-{\rm Tr}[\hat{\sigma}_z \otimes\hat{\sigma}_z \varrho]),
\label{w2fid}
\end{equation}
showing that
only the local measurement settings $\sigma_x^{\otimes 6}$, $\sigma_y^{\otimes 6}$ and $\sigma_z^{\otimes 6}$ are required for an experimental implementation. In fact, any two-qubit correlation ${\rm Tr}[\sigma_k^i \otimes \sigma_k^j \varrho]$ can be obtained from the corresponding data. This puts the method described here on an equal footing with the collective-spin based witness described in Sec.~\ref{collectivebased}, in terms of required experimental effort.

We now investigate how this method copes with noise affecting the class of symmetric Dicke states.
By using Eqs.~(\ref{carlo1}), (\ref{carlo2}) and (\ref{carlo3}), together with Eq.~(\ref{w2fid}), it is straightforward to show that
\begin{equation}
\begin{aligned}
\langle \hat{F}_{\psi^+}\rangle^{ad}&=\frac{1}{2} e^{-2 \gamma } (\alpha_n+e^{\gamma}(1+\alpha_n)-1),\\
\langle \hat{F}_{\psi^+}\rangle^{dp}&=\alpha_n(\gamma-1 )^2-\frac{1}{4}(\gamma-2)\gamma,\\
\langle \hat{F}_{\psi^+}\rangle^{pd}&=\frac{1}{2}(1+e^{-2 \gamma })\alpha_n.
\end{aligned}
\end{equation}
In Fig.~\ref{figRed} {\bf (b)} we show the behavior of $\langle \hat{F}_{\psi^+}\rangle^{ch}$ for the three channels. As soon as $\gamma \ge \text{ln}[(1+\alpha_n+(\alpha_n^2+6\alpha_n-3)^{1/2})/2]$, $\gamma \ge1-(4 \alpha_n-1)^{-1/2} $ and $\gamma \ge -\frac{1}{2} \text{ln}[(1-\alpha_n)/\alpha_n]$ respectively, one finds disconnected sets in an AD-, DP- and PD-affected $\miniket{D^{(n/2)}_n}$ state using $\langle\hat{\cal W}_r \rangle$. These thresholds are larger than those corresponding to the use of a collective-spin based entanglement witness for any value of $n$ that we could quantitatively consider (see discussion in Sec.~\ref{collectivebased}), as a result of the smaller dimension of the states being tested.

\section{State discrimination via characteristic operators}
\label{statediscri}

The variety of ways multipartite entanglement can be shared by an $n$-qubit register requires ways to determine if an experimental state belongs to one of the known classes of entanglement. This can be achieved by relying on the formalism of {\it state discriminators}~\cite{kiesel}, which have been experimentally implemented and used in order to assess the classes of four-qubit entangled states~\cite{kiesel}. Here, we study the reliability of such methods for symmetric Dicke states suffering effects of noisy channels.

Consider multi-qubit operators having $\miniket{D^{(n/2)}_n}$ as a non-degenerate eigenstate associated with the largest possible eigenvalue of the operator's spectrum. We call such operators {characteristic} of the state $\miniket{D^{(n/2)}_n}$. At least one operator having these features exists, {\it i.e.} the fidelity operator $\miniket{D^{(n/2)}_n}\minibra{D^{(n/2)}_n}$. However, this is not the only option and other characteristic operators can be designed, requiring far less local measurement settings than the fidelity one. A systematic approach would require the decomposition of the fidelity operator into tensor products of single-qubit Pauli operators (see Appendix B). Out of them, only the genuine $n$-qubit correlators should be selected to be combined together in a way so as to construct a proper characteristic operator. For instance, for $n=6$ qubits prepared in $\miniket{D^{(3)}_6}$ we can build up the Bell-Mermin operator~\cite{mermin} $\hat{\cal B}_{D^{(3)}_6}\!=\!\sum_{k=x,y}\hat{\cal O}_k/20$ with
$\hat{\cal O}_k\!=\!\hat{\sigma}^1_{k}\otimes[\hat{\sigma}^{\otimes{5}}_k-\sum_{l}\hat{P}_l(\hat{\sigma}^{\otimes{3}}_k\otimes\hat{\sigma}^{\otimes{2}}_z-\hat{\sigma}_k\otimes\hat{\sigma}^{\otimes{4}}_z)]$.
This is characteristic for $\ket{D^{(3)}_6}$, which is an eigenstate with associated eigenvalue $1$ (the maximum within $\hat{\cal B}_{D^{(3)}_6}$'s spectrum). However, if we simply take the negative terms $\hat{\cal D}_{D^{(3)}_6}\!\propto\!-\sum_{k=x,y}\hat{\sigma}^1_{k}\otimes\sum_{l}\hat{P}_{l}(\hat{\sigma}_{k}^{\otimes 3}\otimes\hat{\sigma}^{\otimes{2}}_{z})$ in this expression, we find that $\miniket{D^{(3)}_6}$ is still an eigenstate of maximum eigenvalue (enforced to be $1$ upon renormalization). Representatives of any other class of entanglement will achieve expectation values smaller than $1$. This can be used for effective entanglement-class discrimination. For instance, six-qubit GHZ states transformed by local unitary operations (LU) or stochastic local operations supported by classical communication (SLOCC)~\cite{NIELSEN} yield expectation values no-larger than $0.833$. Thus, if an experimental state $\rho_{exp}$ of $n=6$, thought to be close to the symmetric Dicke family, gives $\langle\hat{\mathcal{D}}_{\rho_{exp}}\rangle>0.833$, one can exclude any GHZ-like character. This can be adapted to any other class of GME states. Here, without affecting the generality of our study, we concentrate on the discrimination between noise-affected symmetric Dicke states and the $n$-qubit GHZ class. We thus construct the streamlined characteristic operators
\begin{equation}
\label{discriminators}
\begin{split}
\hat{\mathcal{D}}_{D^{(n/2)}_n}\!=\!&-{\cal N}\sum_{k=x,y}\hat{\sigma}^1_{k}\otimes\sum_{l}\hat{P}_{l}(\hat{\sigma}_{k}^{\otimes n-3}\otimes\hat{\sigma}^{\otimes{2}}_{z}),
\end{split}
\end{equation}
where ${\cal N}$ is a normalisation factor taken so that the eigenvalue corresponding to $\miniket{{D^{(n/2)}_n}}$ is $1$ and found by noticing that $-\minibra{D^{(n/2)}_{n}}{\hat{\sigma}^{\otimes{n-2}}_k\otimes\hat{\sigma}^{\otimes{2}}_z}\miniket{D^{(n/2)}_{n}}=n/(2n-2)$ for $k=x,y$ and that the number of possible permutations involved in Eq.~(\ref{discriminators}) is $C^{n-3}_{n-1}$, so that ${\cal N}={2}/[n(n-2)]$. We thus see the effects that a channel has on the expectation value of discrimination operators. For AD, PD and DP noise, respectively, are
\begin{equation}
\label{resultsDisc}
\begin{aligned}
\text{Tr}[{\hat{\mathcal{D}}_{D^{(n/2)}_n}\rho_{ad}}] &\!=\!-e^{-\frac{n\gamma }{2}} (e^{\gamma }-2),\\
\text{Tr}[{\hat{\mathcal{D}}_{{D^{(n/2)}_n}} \rho_{pd}}] &\!=\! e^{-(n-2) \gamma },\\
\text{Tr}[{\hat{\mathcal{D}}_{{D^{(n/2)}_n}} \rho_{dp}}] &\!=\!  (1-\gamma )^n.
\end{aligned}
\end{equation}
The first two equations can be understood by using arguments analogous to those valid for collective-spin witness operators. The third identity of Eq.~(\ref{resultsDisc}) is clarified considering the analogy between depolarizing channel and single-qubit white noise.
\begin{figure}[t]
\psfig{figure=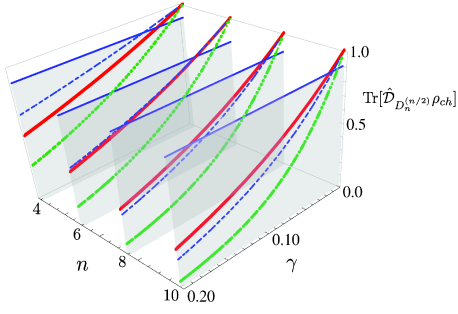,width=6cm,height=4cm}
\caption{(Color online) State discrimination via characteristic operators for symmetric Dicke states with $n=4,6,8,10$. The continuous lines show the behavior of AD-affected $\miniket{D^{(n/2)}_n}$, the dashed lines are for a PD channel, while the lowest dotted lines are for a DP mechanism. Each channel is assumed to be characterized by a rate $\gamma$. The horizontal lines show the maximum expectation values of the characteristic operators $\hat{\cal D}_{D^{(n/2)}_n}$ over SLOCC-equivalent GHZ states of $n$ qubits. The shading highlights the regions where discrimination between symmetric Dicke states and GHZ class is no longer possible. }
\label{fig4}
\end{figure}
In Fig.~\ref{fig4} we show the behavior of Eqs.~(\ref{resultsDisc}) for $n=4,6,8$ and $10$ as the amount of the respective noise-influence increases. We also show the bound associated with the $n$-qubit GHZ class. Thus, the shading in Fig.~\ref{fig4} represents the region where discrimination between one of the channel-affected Dicke states and the GHZ class is not possible. After a certain noise strength, we can no longer determine if the state being studied is a noise-affected Dicke state or a state from the GHZ class. The decay increases more sharply with larger $n$ and the DP channel has the worst effects. We are working on the design of a strategy based on filtering operations which might allow one to gain robustness of this discrimination procedure against noise. However, the potential of this tool is already seen at the level of practical implementability. For instance, the $n=6$ version of Eq.~(\ref{discriminators}) can be decomposed as
$\hat{\mathcal{D}}_{D^{(3)}_6}\!=\!\sum_{k=x,y}\!\hat{\sigma}^1_{k}\!\otimes\![\frac{1}{6}\bigotimes^6_{j=2}(\hat{\sigma}^j_{z}\!+\!\hat{\sigma}^j_{k})\!-\!\frac{1}{6}\bigotimes^6_{j=2}(\hat{\sigma}^j_{z}\!-\!\hat{\sigma}^j_{k})-\!\frac{1}{12}\bigotimes^6_{j=2}(\hat{\sigma}^j_{z}+2\hat{\sigma}^j_{k})\!+\!\frac{1}{12}\bigotimes^6_{j=2}(\hat{\sigma}^j_{z}-2\hat{\sigma}^j_{k})+{5}{}\bigotimes^6_{j=2}\hat{\sigma}^j_{k}]$,
which only requires 10 local measurement settings for its implementation. Finding out the explicit decomposition of discrimination operators for general $n$ is, however, a daunting problem.

\section{Correlation Function}
\label{correlationbased}

In this Section we shift the focus of our discussion from GME to the quantum-coherence properties of the class of states under investigation.
Our approach here considers the expectation value of the $n$-qubit correlation operator
\begin{equation}
\label{defCorrelation}
\hat{\cal C}(\vartheta)\!=\!(\cos\vartheta\hat{\sigma}_k+\sin\vartheta\hat{\sigma}_j)^{\otimes{n}}
\end{equation}
with $k\neq{j}=x,y,z$. Through this, one probes the coherence of each element of the register along a direction (in the single-qubit Bloch sphere) lying in the plane formed by the unit vectors $\bf{k}$ and $\bf{j}$. The use of such a multi-qubit correlator is common in the assessment of the properties of GHZ states for quantum metrology purposes~\cite{blattwineland}: as a result of $n$-qubit coherence, when $k=x$ and $j=y$ the expectation value of Eq.~(\ref{defCorrelation}) oscillates with $\vartheta$ at a frequency that depends on $n$. Here, we shall investigate collective coherence in an $n$-qubit symmetric Dicke state by means of Eq.~(\ref{defCorrelation}). This requires the implementation of a single measurement setting per value of $\vartheta$ and is thus an experimentally favorable tool for multipartite state characterization. For the sake of definiteness, here we concentrate on $k=x$ and $j=z$, although any other choice is equally suitable.

\begin{figure}[t]
\psfig{figure=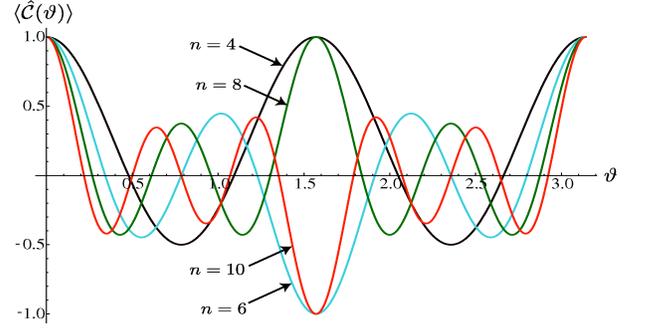,width=8cm,height=4.3cm}
\caption{(Color online) Behavior of the expectation value of the $n$-qubit correlation function calculated over symmetric Dicke states $\miniket{D^{(n/2)}_n}$ with $n=4,6,8$ and $10$. A parity-dependent effect related to the number $n/2$ of excitations in the state being studied is seen: states with an even (odd) number of excitations give rise to a positive (negative) expectation value at $\theta=\pi/2$. Moreover, the position of secondary maxima and minima is $n$-dependent. The beating is an effect of quantum coherence in the state.}
\label{figCorrPure}
\end{figure}

For a pure $\miniket{D^{(n/2)}_n}$ state, the expectation value of  the multi-point correlator is easily found using basic combinatorial arguments and the symmetries in the class of states at hand. We have\begin{equation}
\label{purecaseCorrelation}
\minibra{D^{(n/2)}_n}\hat{\cal C}(\vartheta)\miniket{D^{(n/2)}_n}\!=\!\sum^{n/2}_{k=0}(-1)^k(C^k_{n/2})^2(\cos\vartheta)^{n-2k}(\sin\vartheta)^{2k}.
\end{equation}
In general, Eq.~(\ref{purecaseCorrelation}) exhibits an $n$-dependent oscillatory behavior whose features strongly depend also on the parity of $n/2$, as shown in Fig.~\ref{figCorrPure}. The expression corresponding to $n=6$ has been used in Ref.~\cite{noidicke} in order to contribute to the characterization of $\miniket{D^{(3)}_6}$.

\begin{figure}[t]
{\bf (a)}\hskip3cm{\bf (b)}\hskip3cm{\bf (c)}
\psfig{figure=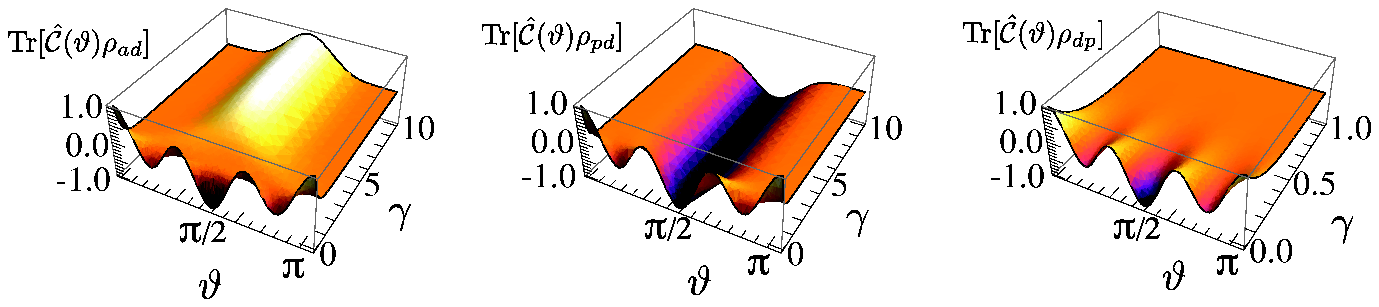,width=8cm,height=2.0cm}\\
\center{{\bf (d)}\hskip2cm{\bf (e)}}
\center{\psfig{figure=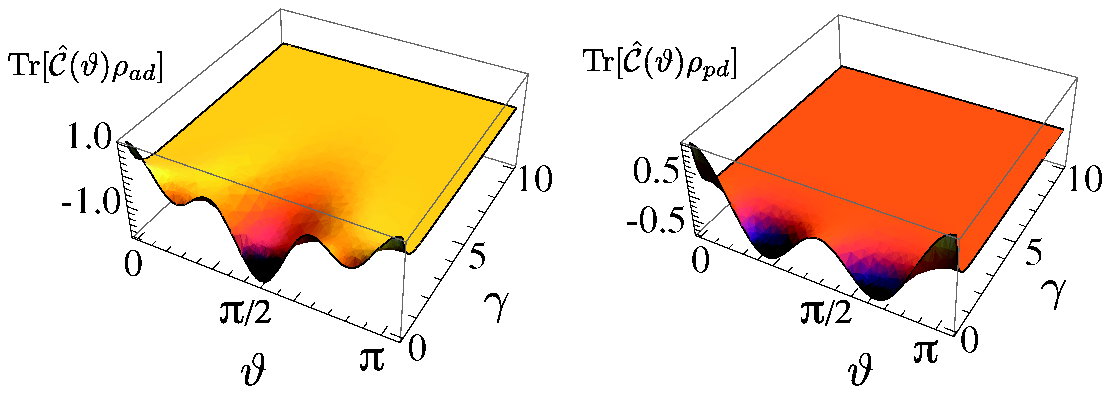,width=5cm,height=2cm}}
\caption{(Color online) We show $\text{Tr}[\hat{\cal C}(\vartheta)\rho_{ad}]$ [panel {\bf (a)}], $\text{Tr}[\hat{\cal C}(\vartheta)\rho_{pd}]$ [panel {\bf (b)}] and $\text{Tr}[\hat{\cal C}(\vartheta)\rho_{dp}]$ [panel {\bf (c)}] against $\vartheta$ and the respective $\gamma$ for a noise affected $\miniket{D^{(3)}_6}$ state. Parity effects associated with the various channels considered are shown. Panels {\bf (d)} and {\bf (e)} show the AD and PD case, respectively, when the classical asymptotic behavior is subtracted.}
\label{figCorr3D}
\end{figure}

Here, we shall study the behavior of $\langle\hat{\cal C}(\vartheta)\rangle$ when a symmetric Dicke state  is subjected to noise effects.  By using again the expressions of the single-qubit Pauli operators transformed upon the action of a given channel (see Sec.~\ref{collectivebased}), one can prove that Eq.~(\ref{purecaseCorrelation}) remains almost invariant under environmental action. The only modification is at the level of oscillation amplitudes, which are changed by a $\gamma$-dependent factor. Explicitly, we have found the following universal form
\begin{equation}
\label{correlationAD}
\text{Tr}[\hat{\cal C}(\vartheta)\rho_{ch}]\!=\!\sum^{n/2}_{k=0}(C^k_{n/2})^2\Gamma^{ch}_{n,k}(\gamma)(\cos\vartheta)^{n-2k}(\sin\vartheta)^{2k}
\end{equation}
with $ch=\{ad,pd,dp\}$ and $\Gamma^{ad}_{n,k}(\gamma)\!=\!e^{-(\frac{n}{2}-k)\gamma}(1-2e^{-\gamma})^k$,
$\Gamma^{pd}_{n,k}(\gamma)\!=\!(-1)^ke^{-({n}-2k)\gamma}$ and $\Gamma^{dp}_{n,k}(\gamma)\!=\!(-1)^k(1-\gamma)^n$ being the channel-specific factors responsible for the loss of coherence in the state. Our claim is that the beating effect responsible for the rich oscillatory structures shown in Fig.~\ref{figCorrPure}  arises only in virtue of the quantum coherences within $\miniket{D^{(n/2)}_n}$. In fact,
while $\Gamma^{dp}_{n,k}(\gamma)\rightarrow{0}$ as $\gamma$ grows, so as to progressively kill any oscillations, $\Gamma^{ad}_{n,n/2}(\gamma)\rightarrow{1}$ and $\Gamma^{pd}_{n,n/2}(\gamma)\rightarrow{(-1)^{n/2}}$, with $\Gamma^{ad,pd}_{n,k}(\gamma)\rightarrow0~\forall{k<n/2}$. This implies that, as the coherences in the $n$-qubit state disappear under an AD or PD channel, the modulus of corresponding correlation functions simply becomes $|(\sin\vartheta)^{n}|$. Thus, although an oscillatory behavior is still kept in these cases, no beating is found. A difference can be found between the trends corresponding to AD and PD channels: a PD channel would progressively destroy the off-diagonal elements of a density matrix without affecting its populations. Asymptotically, this results in a diagonal state whose only non-zero entries are those corresponding to states having $n/2$ $\ket{0}$'s. If we now take $\vartheta=\pi/2 $ in the $n$-qubit correlation operator, we easily understand that its expectation value over the PD-affected state can only be $\pm{1}$, depending on the parity of $n/2$. On the other hand, such a parity-effect is absent in the AD case. Indeed, asymptotically, this channel would reduce any symmetric Dicke state to its collective ground state, which can only give $\langle\hat{\cal C}(\pi/2)\rangle=1,~\forall{n}$. These features are well illustrated in Figs.~\ref{figCorr3D} {\bf (a)}, {\bf (b)} and {\bf (c)} for the case of a symmetric six-qubit Dicke state.

Such an asymptotic classical behavior tends to mask the trend that the beating follows in the changes undergone by the state. We have thus stripped the correlation functions from such contributions by subtracting their Fourier-series expansions from the analogous one of  $\text{Tr}[\hat{\cal C}(\vartheta)\rho_{ad,pd}]$. The results, which highlight the sole decrease in visibility of the fringe of beating induced by quantum correlations, are shown in Figs.~\ref{figCorr3D} {\bf (d)} and {\bf (e)}, which show an evident exponential decay against $\gamma$. Such a procedure is not necessary for the DP, which smoothly flattens the correlation function to zero. The identification of a general trend against the number of qubits in a state is a task made difficult by the $\vartheta$-dependence of such figures of merit: states affected by different channels give rise to maxima and minima of $\text{Tr}[\hat{\cal C}(\vartheta)\rho_{ch}]$ located at different values of the angle $\vartheta$. One can extract useful indicative information by looking, for instance, at the correlations corresponding to $\vartheta=0$, for a set channel and increasing number of qubits, to find that the exponential decay of correlations induced by a growing $\gamma$ becomes faster for larger $n$.
Overall, the analytic expressions provided here for this set of relevant noise channels embody a valuable tool for the experimental characterization of symmetric Dicke states. The theoretical curves can indeed be used to fit the points acquired, experimentally, by tuning the parameters of a setup in a way so as to properly select the direction $\vartheta$ along which one would like to probe quantum coherence~\cite{noidicke}.

\section{Conclusions}
\label{conclusions}

We have studied both GME-detecting and state-characterizing tools for the class of symmetric Dicke states of an arbitrary number of qubits, under the influences of noise. Our investigation starts from standard instruments such as fidelity-based entanglement witnesses and collective-spin operators and unveils some experimentally-friendly ways to improve their resilience to noise. Besides its pragmatic features, which make our study explicitly devoted to the prediction and evaluation of the performances of an experimental setup, we have revealed interesting characteristics of noisy channels and their influence on GME witnesses. For instance, we have clearly shown that resilience to PD channels cannot be achieved by means of simple filtering operations performed on fidelity-based entanglement witnesses.
In light of the steady technological progress in a variety of experimental settings, ranging from linear optics to solid-state systems, we expect our results to be useful for the purposes of a complete characterization of multipartite entangled states generated or evolving in the presence of environmental effects. Our results should also help in the context of reliably determining the quality of experimental symmetric Dickes states for use in quantum networking tasks such as quantum secret sharing and open-destination teleportation.
\acknowledgments

We thank G. Cronenberg, C. Di Franco, M. S. Kim and R. Prevedel for invaluable discussions and encouragement. We acknowledge support from DEL and the UK EPSRC through QIPIRC. MP thanks EPSRC for financial support (EP/G004579/1).

\renewcommand{\theequation}{A-\arabic{equation}}
\setcounter{equation}{0}  
\section*{APPENDIX A}  

In this Appendix, we outline the steps involved in determining the biseparability bound used in the entanglement witness based on collective-spin operators. We start by considering the collective operator
\begin{equation}
\hat{S}_{n}=\hat{J}^2_x+\hat{J}^2_y+\alpha\hat{J}^2_z
\end{equation}
with $\alpha\in\mathbb{R}$. Upon expansion of each collective-spin operator and using the ``one-vs-$(n-1)$" qubit bipartition, we get the expression
\begin{equation}
\hat{S}_{n}=(\frac{n}{2}+\frac{n\alpha}{4})\openone+\frac{1}{2}(\hat{\sigma}^1_{x}\hat{Q}_x+\hat{\sigma}^1_{y}\hat{Q}_y+\alpha\hat{\sigma}^1_{z}\hat{Q}_z+\hat{R}_\alpha),
\end{equation}
where $\hat{Q}_k\!=\!\sum^n_{j=2}\hat{\sigma}^j_{k}$ and $\hat{R}_\alpha\!=\!\hat{R}_x+\hat{R}_y+\alpha\hat{R}_z$ with $\hat{R}_k\!=\!\sum^{n-1}_{j=2}\hat{\sigma}^j_{k}\sum^n_{l=j+1}\hat{\sigma}^l_{k}~(k=x,y,z)$. We take the expectation value of $\hat{S}_n$ over the biseparable state of qubit $1$ and system $2\rightarrow{n}$, so that
\begin{equation}
\begin{aligned}
&\valmed{\hat{S}_n}\!=\!\frac{n}{2}\!+\!\frac{n\alpha}{4}\!+\!\frac{1}{2}[\sum_{k=x,y}\valmed{\hat{\sigma}^1_{k}}\valmed{\hat{Q}_k}\!+\!\alpha\valmed{\hat{\sigma}^1_{z}}\valmed{\hat{Q}_z}+\valmed{\hat{R}_\alpha}]\\
&={\frac{n}{2}\!+\!\frac{n\alpha}{4}}\!+\!\frac{1}{2}\valmed{{\hat{\Omega}}_{\alpha}}\le{\frac{n}{2}+\frac{n\alpha}{4}}+\frac{1}{2}\max_{|r|^2\le{1}}(\lambda_\alpha)=b_{bs}(\alpha)
\end{aligned}
\end{equation}
with $r=(x,y,z)$, $\{\lambda_{\alpha}\}$ the set of eigenvalues of $\hat{\Omega}_{\alpha}$, $x=\valmed{\hat{\sigma}^1_{x}}$ and analogous expressions for $y$ and $z$. The diagonalization of $\hat{\Omega}_\alpha$ can be performed numerically for a few relevant cases, therefore allowing the quantification of the upper bound for biseparability associated with this splitting. For the specific cases of  $n=4,6$ and $8$, the ``one-vs-$(n-1)$" splitting gives the largest bound among all other possible bipartitions, thus providing the threshold $b_{bs}(\alpha)$ for the detection of GME (see Sec.~\ref{collectivebased}). Quantitatively, we have $b_{bs}(0)=5.23,11.018$ and $18.83$ for $n=4,6$ and $8$ respectively.

\renewcommand{\theequation}{B-\arabic{equation}}
\setcounter{equation}{0}  
\section*{APPENDIX B}  

We discuss the local Pauli decomposition of the fidelity operator $\ketbra{D_{n}^{(n/2)}}{D_{n}^{(n/2)}}$ and provide a useful method for reducing the number of local measurement settings experimentally required to measure it. To decompose a given $n$-qubit projector $\ketbra{\phi}{\phi}$ into local Pauli form, we use the relation
\begin{equation}
\ketbra{\phi}{\phi}=\frac{1}{2^{n}}\sum_{i_1,i_2...i_n}C_{i_1,i_2..i_n}(\hat{\sigma}_{i_1} \otimes \hat{\sigma}_{i_2}\otimes..\otimes\hat{\sigma}_{i_n}),
\label{pauli}
\end{equation}
where we have introduced the correlation tensor $C_{i_1,i_2...i_n}=\bra{\phi}\hat{\sigma}_{i_1} \otimes \hat{\sigma}_{i_2}\otimes..\otimes\hat{\sigma}_{i_n}\ket{\phi}$ and $i_n \in \{0,x,y,z\}$, with $\hat{\sigma}_0=\hat{\openone}$. Using Eq.~(\ref{pauli}), the correlation tensor corresponding to $|D^{(2)}_4\rangle\langle{D}^{(2)}_4|$ has 40 non-zero elements
\begin{equation}
\label{D4}
\begin{aligned}
&\ketbra{D^{(2)}_4}{D^{(2)}_4}\!=\!\frac{1}{16}(\openone^{\otimes 4}\!+\!\sum_{k=x,y,z}\hat{\sigma}_k^{\otimes 4}
+\frac{1}{3}\sum_{\pi}[\sigma_x^{\otimes 2}\sigma_y^{\otimes 2}\\
&+2(\openone^{\otimes 2}\hat{\sigma}_x^{\otimes 2}\!+\!\openone^{\otimes 2}\sigma_y^{\otimes 2}\!+\!\openone^{\otimes 2}\sigma_z^{\otimes 2}\!-\!\sigma_x^{\otimes 2}\sigma_z^{\otimes 2}\!-\!\sigma_y^{\otimes 2}\sigma_z^{\otimes 2})] ),
\end{aligned}
\end{equation}
where $\sum_{\pi}$ indicates all distinct permutations of the operators. However, such a 40-element decomposition can be significantly streamlined by using the relation (valid for $i,j=x,y,z)$
\begin{equation}
\sum_{\pi}\sigma_i^{\otimes 2}\sigma_j^{\otimes 2}\!=\!\frac{1}{2}[(\sigma_i+\sigma_j)^{\otimes 4}+(\sigma_i-\sigma_j)^{\otimes 4}]-\sigma_i^{\otimes 4}-\sigma_j^{\otimes 4},
\label{comp}
\end{equation}
through which one can rewrite the $\sum_{\pi}$ term of Eq.~(\ref{D4}) in terms of the 9 local measurement settings:$\sigma_{x,y,z}^{\otimes 4}$,\,$[(\sigma_x\!\pm\!\sigma_{y,z})/\sqrt{2}]^{\otimes 4}$ and $[(\sigma_y\!\pm\!\sigma_{z})/\sqrt{2}]^{\otimes 4}$. Thus, the symmetries present in the state's decomposition have been exploited in a way so as to reduce the number of measurements. Using the compacting techniques outlined here for $|D_{4}^{(2)}\rangle$, one should be able to apply them to arbitrary sized symmetric Dicke states $|D_{n}^{(n/2)}\rangle$ to obtain significant reductions in the number of local measurement settings.

For example, using Eq.~(\ref{pauli}), one finds that the correlation tensor corresponding to $|D^{(3)}_6\rangle\langle{D}^{(3)}_6|$ has 544 non-zero elements
\begin{equation}
\begin{aligned}
&\ketbra{D^{(3)}_6}{D^{(3)}_6}\!=\!\sum_{\pi}(\frac{1}{320}[\sigma_x^{\otimes 2}\sigma_y^{\otimes 4}+\sigma_x^{\otimes 4}\sigma_y^{\otimes 2}-\sigma_x^{\otimes 2}\sigma_y^{\otimes 2}\sigma_z^{\otimes 2} \\
&+\openone^{\otimes 2}\sigma_x^{\otimes 2}\sigma_y^{\otimes 2}+\openone^{\otimes 2}\sigma_z^{\otimes 4}-\openone^{\otimes 4}\sigma_z^{\otimes 2} -\openone^{\otimes 2}\sigma_x^{\otimes 2}\sigma_z^{\otimes 2}\\
&-\openone^{\otimes 2}\sigma_y^{\otimes 2}\sigma_z^{\otimes 2}] +\frac{3}{320}[\sigma_x^{\otimes 2}\sigma_z^{\otimes 4}-\sigma_y^{\otimes 2}\sigma_z^{\otimes 4}-\sigma_x^{\otimes 4}\sigma_z^{\otimes 2}\\
&-\sigma_y^{\otimes 4}\sigma_z^{\otimes 2}+\openone^{\otimes 4}\sigma_x^{\otimes 2}+\openone^{\otimes 2}\sigma_x^{\otimes 4}+\openone^{\otimes 4}\sigma_y^{\otimes 2}+\openone^{\otimes 2}\sigma_y^{\otimes 4}])\\
&+\frac{1}{64}[\sigma_x^{\otimes 6}+\sigma_y^{\otimes 6}+\openone^{\otimes 6}-\sigma_z^{\otimes 6}].
\label{D6}
\end{aligned}
\end{equation}
Using Eq.~(\ref{comp}), together with the relations
\begin{equation}
\begin{aligned}
&\sum_{\pi}\sigma_x^{\otimes 2}\sigma_y^{\otimes 2}\sigma_z^{\otimes 2}=\frac{1}{4}[(\sigma_x+\sigma_y+\sigma_z)^{\otimes 6}+(\sigma_x+\sigma_y-\sigma_z)^{\otimes 6} \\
&\!+\!(\sigma_x\!-\!\sigma_y\!+\!\sigma_z)^{\otimes 6}\!+\!(\sigma_x\!-\!\sigma_y\!-\!\sigma_z)^{\otimes 6}]\!
-\!\sum_{\pi}(\sigma_x^{\otimes 2}\sigma_y^{\otimes 4}\!+\!\sigma_x^{\otimes 4}\sigma_y^{\otimes 2})\\
&\!-\!\sum_{\pi}\sum_{k=x,y}(\sigma_k^{\otimes 2}\sigma_z^{\otimes 4}+\sigma_k^{\otimes 4}\sigma_z^{\otimes 2})
\!-\!\sum_{k=x,y,z}\sigma_k^{\otimes 6},\\
&\sum_{\pi}(\sigma_i^{\otimes 2}\sigma_j^{\otimes 4}+\sigma_i^{\otimes 4}\sigma_j^{\otimes 2})\!=\!\frac{1}{2}[(\sigma_i\!+\!\sigma_j)^{\otimes 6}\!+\!(\sigma_i-\sigma_j)^{\otimes 6}]\\
&-\sigma_i^{\otimes 6}-\sigma_j^{\otimes 6},\\
&\sum_{\pi}(\sigma_i^{\otimes 2}\sigma_j^{\otimes 4}-\sigma_i^{\otimes 4}\sigma_j^{\otimes 2})\!=\!\frac{1}{24}\{(\sigma_i+2\sigma_j)^{\otimes 6}+(\sigma_i-2\sigma_j)^{\otimes 6}\\
&-4[(\sigma_i+\sigma_j)^{\otimes 6}+(\sigma_i-\sigma_j)^{\otimes 6}]-10\sigma_i^{\otimes 6}-136\sigma_j^{\otimes 6}\},
\end{aligned}
\end{equation}
one can rewrite the permutations of Eq.~(\ref{D6}) in terms of the following 21 local measurement settings: $\sigma_{x,y,z}^{\otimes 6}$, $[(\sigma_x\pm\sigma_{y,z})/\sqrt{2}]^{\otimes 6}$, $[(\sigma_y\pm\sigma_{z})/\sqrt{2}]^{\otimes 6}$, $[(\sigma_{x,y}\pm2\sigma_z)/\sqrt{5}]^{\otimes 6}$, $[(\sigma_z\pm2\sigma_{x,y})/\sqrt{5}]^{\otimes 6}$ and $[(\sigma_x\pm\sigma_y\!\pm\!\sigma_z)/\sqrt{3}]^{\otimes 6}$. Such a reduction provides a huge advantage for measuring the fidelity of $|D_{6}^{(3)}\rangle$ in a given experimental setup.


\begin{thebibliography}{99}
\bibitem{footnote0} Our results do not depend on the number of qubits, which makes them well-suited for application to any experimental setup able to generate GME of the class we consider.
\end{thebibliography}
\end{document}